%% file: main.tex
\documentclass[sigconf,natbib]{acmart} 

\AtBeginDocument{%
  \providecommand\BibTeX{{%
    \normalfont B\kern-0.5em{\scshape i\kern-0.25em b}\kern-0.8em\TeX}}}

\usepackage{enumitem}
\usepackage{multicol}
\usepackage{multirow}
\usepackage{soul}
\setul{0.3ex}{}
\usepackage{makecell}
\usepackage{colortbl}
\usepackage{xspace}
\usepackage{graphicx}
\usepackage{subfigure}
\usepackage{balance}

\setcopyright{acmcopyright}
\copyrightyear{2023}
\acmYear{2023}
\acmDOI{XXXXXXX.XXXXXXX}

\acmPrice{15.00}
\acmISBN{978-1-4503-XXXX-X/18/06}

\acmConference[KDAH CIKM '23]{Fifth Workshop on Knowledge-driven Analytics and Systems impacting Human Quality of Life at CIKM 2023}{October 22,
  2023}{Birmingham, UK}

\acmBooktitle{Fifth Workshop on Knowledge-driven Analytics and Systems impacting Human Quality of Life (KDAH CIKM '23), October 22, 2023, Birmingham, United Kingdom}

\usepackage{xspace}
\newcommand{\modelname}{SPCS\xspace}
\newcommand{\industry}{Meituan\xspace}
\newcommand{\std}[1]{{\scriptsize{$\pm$#1}}}
\newtheorem{mydef}{Definition}
\input{04other/math_commands.tex}
\begin{document}

\title{Modeling Spatiotemporal Periodicity and Collaborative Signal for Local-Life Service Recommendation}



\author{Huixuan Chi$^{1\ast}$, Hao Xu$^{2\ast}$, Mengya Liu$^3$, Yuanchen Bei$^4$, Sheng Zhou$^4$, Danyang Liu$^2$, Mengdi Zhang$^2$}\thanks{* Both authors contributed equally.}

\affiliation{\vspace{1.5ex}
$^1$Institute of Computing Technology, Chinese Academy of Sciences \country{China} \\
$^2$Meituan, China \quad
$^3$University of Southampton, UK \\ 
$^4$Zhejiang University, China \\
\vspace{0ex} 
chihuixuan21s@ict.ac.cn, kingsleyhsu1@gmail.com\\ 
}

\renewcommand{\shortauthors}{Huixuan Chi, Hao Xu, Mengya Liu, Yuanchen Bei, Sheng Zhou, Danyang Liu, \& Mengdi Zhang}

\ccsdesc[500]{Information systems~Location based services}

\keywords{Spatiotemporal Periodicity, Collaborative Signal, Recommendation}


\begin{abstract}

\input{01tex/000abstract.tex}

\end{abstract}

\maketitle

\section{Introduction}

\input{01tex/012introduction.tex}

\label{sec:intro}



\section{Related Works}

\input{01tex/050related.tex}
\label{sec:related}

\section{Methodology}

\input{01tex/030method.tex}
\label{sec:method}


\section{Experiment}

\input{01tex/040experiment.tex}

\label{sec:experiment}


\section{Conclusion}

\input{01tex/060conclusion.tex}
\label{sec:conclusion}


\bibliographystyle{03ref/ACM-Reference-Format}
\balance
\bibliography{03ref/reference}


\end{document}

%% file: 04other/math_commands.tex
\usepackage{amsmath,amsfonts,bm}

\def\eqref#1{equation~\ref{#1}}

\def\1{\bm{1}}

\DeclareMathAlphabet{\mathsfit}{\encodingdefault}{\sfdefault}{m}{sl}
\SetMathAlphabet{\mathsfit}{bold}{\encodingdefault}{\sfdefault}{bx}{n}

%% file: 01tex/000abstract.tex
Online local-life service platforms provide services like nearby daily essentials and food delivery for hundreds of millions of users. 
Different from other types of recommender systems, local-life service recommendation has the following characteristics:
(1) \textit{spatiotemporal periodicity}, which means a user's preferences for items vary from different locations at different times.
(2) \textit{spatiotemporal collaborative signal}, which indicates similar users have similar preferences at specific locations and times. 
However, most existing methods either focus on merely the spatiotemporal contexts in sequences, or model the user-item interactions without spatiotemporal contexts in graphs.
To address this issue, we design a new method named \modelname in this paper.
Specifically, we propose a novel \textit{spatiotemporal graph transformer} (SGT) layer, which explicitly encodes relative spatiotemporal contexts, and aggregates the information from multi-hop neighbors to unify spatiotemporal periodicity and collaborative signal.
With extensive experiments on both public and industrial datasets, this paper validates the state-of-the-art performance of \modelname.

%% file: 01tex/012introduction.tex
\begin{figure}[t]
    \centering
    \includegraphics[width=\linewidth]{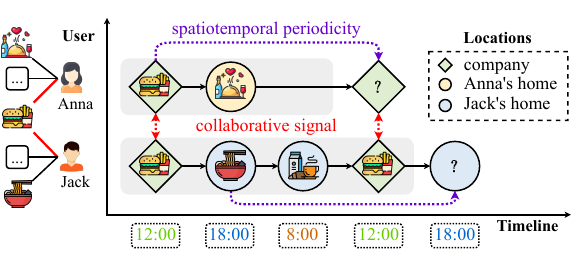}
    \caption{The motivation example. For Anna or Jack, spatiotemporal periodicity is shown in her/his trajectory at different locations and times. For Anna and Jack, spatiotemporal collaborative signal is reflected in their similar preferences at specific locations and times.}
    \label{fig:intro-spatial}
\end{figure}

Online local-life service platforms, such as Meituan\footnote{\url{http://i.meituan.com/}} and Uber Eats\footnote{\url{https://www.ubereats.com/}}, provide services like nearby daily essentials and food delivery for hundreds of millions of users.
Different from other types of recommender systems \cite{zhou2018deep,zhou2019deep}, local-life service recommendation has the following characteristics:
(1) \textit{spatiotemporal periodicity}, which means a user's preferences for items vary from different locations at different times. 
According to the trajectory in Figure \ref{fig:intro-spatial}, Jack enjoys burgers at the company at noon, but he prefers noodles at home in the evening.
(2) \textit{spatiotemporal collaborative signal}, which means similar users exhibit similar preferences at specific locations and times. 
For instance, Anna and Jack share similar preferences while they are at the same company at noon.
However, their preferences diverge in the evening due to their different home locations.
Moreover, due to the strong coupling between these characteristics, it is crucial to directly incorporate spatiotemporal information rather than separately considering spatial or temporal aspects.
Overall, both two characteristics (purple arrow and red arrow in Figure \ref{fig:intro-spatial}) have important implications for local-life service recommendation yet have rarely been studied in existing works. 
In this paper, we focus on simultaneously modeling the spatiotemporal periodicity and collaborative signal in local-life service platforms.

Generally speaking, there are two relevant lines of research for local-life service recommendation as shown in \textbf{Table \ref{tab:compare}}.
One is the sequential-based recommender. 
The majority of studies \cite{liu2016predicting,sun2020go,luo2021stan,lian2020geography,cui2021st} apply recurrent neural networks with spatiotemporal contexts between current and future steps to capture the transitional regularities in sequences. 
However, they ignore the \textit{spatiotemporal collaborative signal} and fail to capture preferences from similar users.
Another line of research is the graph-based recommender. 
Earlier works \cite{wang2019neural,he2020lightgcn} simply adopt GCN layers on the user-item interaction graph for collaborative filtering, ignoring the spatiotemporal contexts.
Recent works \cite{ma2018point,fan2021continuoustime,rossi2020temporal} aim to utilize spatial or temporal information in graphs.
However, these methods either focus solely on the temporal information, or consider merely spatial information (location), which fail to model the \textit{spatiotemporal periodicity}.
In short, existing solutions leave a blank space in simultaneously modeling the spatiotemporal periodicity and collaborative signal for local-life service recommendation.

To address this issue, we propose a novel method, Modeling \textbf{S}patiotemporal \textbf{P}eriodicity and \textbf{C}ollaborative \textbf{S}ignal (\modelname) for local-life service recommendation.
First, to capture the spatiotemporal contexts in graphs, we design the encoding layer, which explicitly encodes relative time interval as \textit{temporal encoding} and relative spatial distance as \textit{spatial encoding}.
Second, to simultaneously model the spatiotemporal periodicity and collaborative signal, we design a novel \textit{spatiotemporal graph transformer} (SGT) layer, which utilizes two encodings and aggregates the spatiotemporal information from multi-hop neighbors.
The main contributions of this paper are as follows:
\begin{itemize}[leftmargin=*]
    \item We propose a novel \modelname method to simultaneously model the spatiotemporal periodicity and collaborative signal for local-life service recommendation. 
    Specifically, we capture the spatiotemporal contexts in the encoding layer, and then aggregate this information from multi-hop neighbors in the spatiotemporal graph transformer (STG) layer.
    \item We conduct extensive experiments on both real-world public and industrial datasets. Experimental results demonstrate the effectiveness of our proposed \modelname.
\end{itemize}

\begin{table}[t]
\caption{Comparison of related methods.}
\label{tab:compare}
\resizebox{\linewidth}{!}{
\begin{tabular}{c|c|c|c|c}
\hline
\multicolumn{1}{c|}{\textbf{Type}} & \multicolumn{1}{c|}{\textbf{Model}} & \multicolumn{1}{c|}{\textbf{Collaborative}} & \multicolumn{1}{c|}{\textbf{Temporal Info}} & \multicolumn{1}{c}{\textbf{Spatial Info}} \\ \hline
\multirow{4}{*}{Sequential-based}               & ST-RNN & $\times$  & \checkmark   & \checkmark     \\
                               & STAN & $\times$  & \checkmark   & \checkmark     \\
                               & TiSASRec & $\times$ & \checkmark & $\times$  \\
                               & SLRC & $\times$ & \checkmark & $\times$  \\ \hline
\multirow{4}{*}{Graph-based}                   & LightGCN & \checkmark& $\times$ & $\times$ \\
                               & SAE-NAD & \checkmark & $\times$ & \checkmark   \\
                               & TGSRec & \checkmark & \checkmark   & $\times$     \\
                               & TGN                                 & \checkmark                                                  & \checkmark                                          & $\times$                                         \\ \hline
                            Our method   & \textbf{\modelname}                          & \checkmark                                                  & \checkmark                                          & \checkmark                                 \\ \hline
\end{tabular}
}
\vspace{-0.1in}
\end{table}

%% file: 01tex/050related.tex

\textbf{Sequential-based Recommenders.} 
Most of the earlier works like GRU4Rec \cite{wu2019sessionbased} and TiSASRec \cite{li2020time} adopt sequential methods, such as RNN \cite{medsker2001recurrent} and self-attention \cite{vaswani2017attention}, to capture the temporal evolution of user preferences.
Besides, some other works \cite{liu2016predicting,cui2021st,luo2021stan} like ST-RNN \cite{liu2016predicting} and STAN \cite{luo2021stan} aim to utilize spatiotemporal contexts between current and future steps to capture the transitional regularities.
However, these methods ignore the \textit{spatiotemporal collaborative signal}, which is essential to capture preferences from similar users at specific locations and times.


\noindent\textbf{Graph-based Recommenders.} 
Early works like NGCF \cite{wang2019neural} and LightGCN \cite{he2020lightgcn} simply adopts GCN layers on the user-item interaction graph without using spatiotemporal contexts.
Later works \cite{ma2018point,han2020stgcn,yang2022getnext} exploit the extensive collaborative signals for item-item relations with location information.
Some other works \cite{wang2019modeling,fan2021continuoustime,rossi2020temporal,tian2022temporal,chi2022long,bei2023cpdg} like TGSRec \cite{fan2021continuoustime} and TGN \cite{rossi2020temporal} attempt different utilization of time intervals with time-decaying functions to capture the temporal collaborative signal.
However, these methods fail to fully utilize the spatiotemporal contexts and ignore the \textit{spatiotemporal periodicity}.
Therefore, we aim to simultaneously model the spatiotemporal periodicity and collaborative signal for local-life service recommendation.

%% file: 01tex/030method.tex
\begin{figure}[t]
    \centering
    \includegraphics[width=\linewidth]{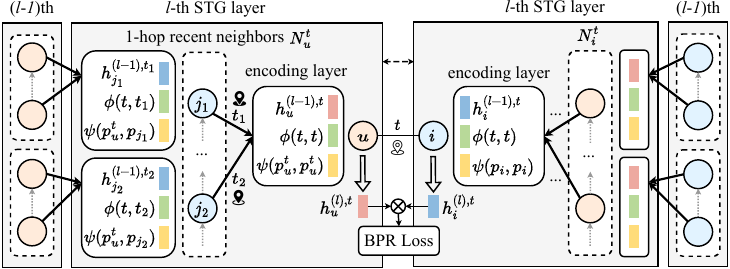}
    \vspace{-0.2in}
    \caption{The overall framework of our proposed \modelname.}
    \label{fig:method-framework}
    \vspace{-0.1in}
\end{figure}

In this section, we first present related terms and formalize the problem of local-life service recommendation, then introduce the three main parts of our \modelname: 
(1) \textit{encoding layer}, which explicitly encodes the spatiotemporal contexts as temporal encoding and spatial encoding.  
(2) \textit{spatiotemporal graph transformer layer}, which combines two encodings and aggregates the information from multi-hop neighbors. 
(3) the prediction and optimization. 
Figure \ref{fig:method-framework} illustrates the overall structure of \modelname.

\subsection{Problem Formulation}

\begin{mydef}
    \textbf{Spatiotemporal User-Item Graph} is defined as $\mathcal{G} = \{\mathcal{U}, \mathcal{I}, \mathcal{E}\}$, where $\mathcal{U}$ and $\mathcal{I}$ is the set of users and items, $\mathcal{E}$ is a set of spatiotemporal edges. Each edge $e_{u, i}^{t} \in \mathcal{E}$ is denoted as a quintuple, $e_{u, i}^{t} = (u, i, t, p_{u}^t, p_i)$, where $u \in \mathcal{U}$, $i \in \mathcal{I}$, $t \in \mathbb{R}^+$, $\{p_{u}^t, p_i\} \subset \mathcal{P}$, and $\mathcal{P}$ is a set of locations with latitude and longitude. Each $(u, i, t, p_{u}^t, p_i)$ means that a user $u$ will interact with item $i$ at time $t$, traveling from location $p_{u}^t$ to location $p_{i}$.
    We denote the neighbor set for user $u$ in the time interval $[0, t]$ as $\mathcal{N}_{u}^t = \{(j, t_j, p_j) \ | \ e_{u, j}^{t_j} \in \mathcal{E}, t_j < t\}$. 
\end{mydef}

\begin{mydef}
    \textbf{Local-Life Service Recommendation.} Given a spatiotemporal user-item graph $\mathcal{G}$ and a new query tuple $(u, t, p_{u}^t)$, local-life service recommendation aims to recommend item list that user $u$ would be interested at time $t$ and location $p_u^t$. 
\end{mydef}

\subsection{Encoding Layer}\label{sec:periodic-encoding}

\subsubsection{\textbf{User/Item Embedding}}

To maintain each user’s (item’s) history in a compressed format, we adopt a widely-used memory mechanism \cite{rossi2020temporal}. 
First, the memory state for a user $u$ (item $i$) at time $t$ is denoted as $\bm{s}_{u}^t$ ($\bm{s}_{i}^t$) $\in \mathbb{R}^c$, initialized as an all-zero vector and updated by the memory mechanism. 
Then, we define $\bm{h}_u^{(\ell - 1), t}$ as the hidden embedding that serves as the input to the $\ell$-th layer for user $u$ at time $t$ (the same applies to the hidden embedding for item $i$).
Note that, in the first layer, $\bm{h}_u^{(0), t} = s_u^t$. 
When $\ell > 1$, it is generated from the previous layer.

\subsubsection{\textbf{Temporal Encoding (TE)}}

Inspired by recent works \cite{xu2020inductive, fan2021continuoustime}, the periodicity can be reflected by relative time intervals between the user's different interactions. 
We design the temporal encoding function as $\phi(\cdot, \cdot) \rightarrow \mathbb{R}^{c_t}$ based on Bochner’s Theorem \cite{loomis2013introduction} to encode the time intervals into embedding explicitly.
Specifically, given two time points $t_1$ and $t_2$, we implement $\phi(\cdot, \cdot)$ as:
\begin{equation}
    \phi(t_1, t_2) = \left[\cos(\omega_1 |t_1 - t_2| + b_1), \cdots, \cos(\omega_{c_t} |t_1 - t_2| + b_{c_t})\right]
\end{equation}
where $\cos(\cdot)$ is the cosine function. $\bm{\omega} = [\omega_1, \cdots, \omega_{c_t}]$ and $\bm{b} = [b_1, \cdots, b_{c_t}]$ are learnable weights and bias of linear transformation for the time interval. 

\subsubsection{\textbf{Spatial Encoding (SE)}}
To quantify the change of locations between different users and items, we design the spatial encoding function as  $\psi(\cdot, \cdot) \rightarrow \mathbb{R}$ to derive the geographical weight.
Specifically, given two locations $p_1$ and $p_2$, we implement $\psi(p_1, p_2)$ as: 
\begin{equation}
    \psi(p_1, p_2) = \frac{1}{f\left(\mathrm{Haversine}(p_1, p_2) / \tau\right) + 1},
\end{equation}
where $\mathrm{Haversine}(\cdot, \cdot)$ denotes the Haversine formula \cite{robusto1957cosine}, which is widely used to calculate the distance from latitude and longitude. 
$\tau$ is used to control the decay rate of the weight.
And $f(\cdot)$ is a mapping function, such as an identity or exponential function.
Note that other effective spatial encoding can also be explored and used in future work.

\subsection{Spatiotemporal Graph Transformer Layer}

In this section, we will introduce the spatiotemporal graph transformer (SGT) layer in two parts:
(1) construction of query, key, and value; (2) spatiotemporal self-attention.
In the following, we take the calculation for user $u$ at time $t$ as an example.



\subsubsection{\textbf{Construction of Query, Key and Value}}

To unify the spatiotemporal information and collaborative signal, we construct the input information of each SGT layer as the combination of hidden node embeddings, temporal encoding, and spatial encoding. 
Specifically, given the query user $(u, t, p_u^t)$, we construct the query vector $\bm{q}^{(\ell), t}$ for user $u$ itself at $\ell$-th layer as:
\begin{equation}\label{equ:query}
    \bm{q}^{(\ell), t} = \left[~\bm{h}_u^{(\ell-1), t} \ \oplus \ \phi(t, t) \ \otimes \ \psi(p_{u}^t, p_{u}^{t})\right], 
\end{equation}
where the layer number $\ell = \{1, \cdots,L\}$. 
And $\oplus$ is the vector concatenate operation and $\otimes$ is the product operation in this paper.
Other operations including summation are possible.
In addition to query user $u$ itself, we also propagate spatiotemporal collaborative information from its neighbors.
We sample $M$ most recent neighbors $(j_1, t_1, p_{j_1}), (j_2, t_2, p_{j_2}), \cdots$ of user $u$ from $\mathcal{N}_{u}^t$.
Similar to the construction of the query vector, the key/value matrix for user $u$'s most recent neighbors can be formulated as: 
\begin{equation}\label{equ:key-value}
    \mathbf{K}^{(\ell), t} = \mathbf{V}^{(\ell), t} = \begin{bmatrix}
    ~ \bm{h}_{j_1}^{(\ell-1), t_1} \ \oplus \ \phi(t, t_1) \ \otimes \ \psi(p_{u}^t, p_{j_1}), \\
    ~ \bm{h}_{j_2}^{(\ell-1), t_2} \ \oplus \ \phi(t, t_2) \ \otimes \ \psi(p_{u}^t, p_{j_2}), \\
    \cdots 
    \end{bmatrix}.
\end{equation}

\subsubsection{\textbf{Spatiotemporal Self-Attention}}

Then, we adopt a self-attention mechanism to propagate the information as follows:
\begin{equation}\label{equ:attn}
    \bm{h}_u^{(\ell), t} = \left(\mathbf{W}_v^{(\ell)} \mathbf{V}^{(\ell), t}\right) \cdot \sigma\left(\frac{[\mathbf{W}_k^{(\ell)} \mathbf{K}^{(\ell), t}]^{\top} [\mathbf{W}_q^{(\ell)} \bm{q}^{(\ell), t}]}{\sqrt{c + c_{t}}}\right),
\end{equation}
where $\sigma(\cdot)$ is the softmax function. 
$\mathbf{W}_{q}^{(\ell)}, \mathbf{W}_{k}^{(\ell)}, \mathbf{W}_{v}^{(\ell)}  \in \mathbb{R}^{c\times (c+c_t)}$ are learnable transformation matrices at $\ell$-th layer. 

By stacking $L$ layers, we can obtain the final embedding for user $u$ as $\bm{h}_{u}^t = \bm{h}_{u}^{(L),t}$. Analogously, for item $i$, we need to alternate the user information to item information, and change the neighbor information in Eq. (\ref{equ:query}), (\ref{equ:key-value}) and (\ref{equ:attn}) according to user-item pairs. Thus, $\bm{h}_{i}^t$ for item $i$ can also be calculated. 
The time complexity of the proposed \modelname is $O\left(|\mathcal{U} \cup \mathcal{I}| \cdot L \cdot \left(M  (c + c_t)  c + Mc \right)\right)$.

\subsection{Prediction and Optimization}

For each $(u, i, t, p_u^t, p_i)$, we can calculate the affinity score $y_{u, i}^t = \mathrm{MLP} (\bm{h}_{u}^t \| \bm{h}_{i}^t)$ between user $u$ and item $i$ at time $t$.
To optimize parameters, we utilize the widely-used pairwise BPR loss \cite{rendle2012bpr} for top-K recommendation, which can be formulated as:
\begin{equation}
    \mathcal{L} = \sum_{u \in \mathcal{U}} \sum_{(u, i, i', t) \in \mathcal{O}} -\ln \sigma \left(y_{u, i}^t - y_{u, i'}^t\right) + \lambda \| \bm{\theta} \|_2^2,
\end{equation}
where $\mathcal{O} = \{(u, i, i', t) \ | \ e_{u, i}^t \in \mathcal{E} , e_{u, i'}^{t} \notin \mathcal{E}\}$ denotes the pairwise training data, and $\sigma(\cdot)$ is the sigmoid function. $ \lambda \| \bm{\theta} \|_2^2$ denotes the $L_2$ regularization for addressing over-fitting. 

%% file: 01tex/040experiment.tex
\subsection{Experimental Settings}

\begin{table}[t]
\caption{Statistics of Datasets.}
\label{tab:datasets}
\resizebox{\linewidth}{!}{
\begin{tabular}{ccccc}
\hline
\multicolumn{1}{c}{\textbf{Dataset}} & \multicolumn{1}{c}{\textbf{\#User}} & \multicolumn{1}{c}{\textbf{\#Item}} & \multicolumn{1}{c}{\textbf{\#Instance}} & \textbf{Timespan}           \\ 
\hline
Gowalla-Food                         & 15,058                               & 26,594                                & 553,121                                  & 2009.1.21-2010.3.11 \\
\industry                              & 17,862                               & 26,236                                & 649,101                                  & 2022.2.14-2022.3.28 \\ 
\hline
\end{tabular}
}
\end{table}

\begin{table*}[htp]
\caption{Comparison results on two datasets. \ul{Underline} means the best baseline, and \textbf{bold} means the best performance. We report the average and standard deviation over 3 independent runs.}
\vspace{-0.05in}
\label{tab:all}
\begin{tabular}{@{}c|cccc|cccc@{}}
\hline
\multicolumn{1}{c|}{\multirow{2}{*}{\textbf{Model}}} & \multicolumn{4}{c|}{\textbf{Gowalla-Food}}                                                                                                                                            & \multicolumn{4}{c}{\textbf{\industry}}                                                                                                                                                 \\ 
\cline{2-5} \cline{6-9}
\multicolumn{1}{c|}{}                                & \multicolumn{1}{c}{\textbf{Hit@10}}     & \multicolumn{1}{c}{\textbf{Hit@20}}     & \multicolumn{1}{c}{\textbf{NDCG@10}}       & \textbf{NDCG@20}                            & \multicolumn{1}{c}{\textbf{Hit@10}}     & \multicolumn{1}{c}{\textbf{Hit@20}}     & \multicolumn{1}{c}{\textbf{NDCG@10}}       & \textbf{NDCG@20}                           \\ 
\hline
ST-RNN                                                & \multicolumn{1}{c}{0.0264\std{0.0002}}          & \multicolumn{1}{c}{0.0397\std{0.0001}}          & \multicolumn{1}{c}{0.0189\std{0.0001}}          & 0.0222\std{0.0001}                               & \multicolumn{1}{c}{0.0185\std{0.0002}}          & \multicolumn{1}{c}{0.0357\std{0.0003}}          & \multicolumn{1}{c}{0.0098\std{0.0001}}          & 0.0131\std{0.0001}                              \\
STAN & 0.1971\std{0.0021} & 0.2459\std{0.0025} & 0.1443\std{0.0021} & 0.1563\std{0.0023} & {0.0846}\std{0.0008} & {0.1351}\std{0.0011} & {0.0421}\std{0.0008} & 0.0485\std{0.0011} \\
TiSASRec & 0.0396\std{0.0017} & 0.0630\std{0.0023} & 0.0214\std{0.0011} & 0.0272\std{0.0012} & 0.0793\std{0.0007} & 0.1276\std{0.0001} & 0.0409\std{0.0001} & 0.0530\std{0.0001} \\ 
SLRC & 0.1837\std{0.0014} & 0.2262\std{0.0008} & 0.1390\std{0.0012} & 0.1441\std{0.0015} & \ul{0.1053}\std{0.0010} & \ul{0.1650}\std{0.0016} & \ul{0.0516}\std{0.0011} & 0.0580\std{0.0011} \\
\hline
LightGCN                                             & \multicolumn{1}{c}{0.0337\std{0.0001}}          & \multicolumn{1}{c}{0.0562\std{0.0003}}          & \multicolumn{1}{c}{0.0109\std{0.0001}}          & 0.0142\std{0.0001}                               & \multicolumn{1}{c}{0.0380\std{0.0002}}          & \multicolumn{1}{c}{0.0646\std{0.0002}}          & \multicolumn{1}{c}{0.0179\std{0.0001}}          & 0.0205\std{0.0001}                              \\
SAE-NAD                                              & \multicolumn{1}{c}{0.0873\std{0.0043}}          & \multicolumn{1}{c}{0.1314\std{0.0049}}          & \multicolumn{1}{c}{0.0541\std{0.0019}}          & 0.0678\std{0.0020}                               & \multicolumn{1}{c}{0.0555\std{0.0013}}          & \multicolumn{1}{c}{0.0938\std{0.0022}}          & \multicolumn{1}{c}{0.0379\std{0.0008}}          & 0.0512\std{0.0012}                              \\
TGSRec & 0.1595\std{0.0163}  & 0.2141\std{0.0116} & 0.1071\std{0.0205} & 0.1208\std{0.0194} & 0.0619\std{0.0007} & 0.0998\std{0.0010} & 0.0315\std{0.0006} & 0.0409\std{0.0006} \\
TGN                                                  & \multicolumn{1}{c}{{\ul{0.2440}}\std{0.0046}}    & \multicolumn{1}{c}{{\ul{0.3041}\std{0.0029}}}    & \multicolumn{1}{c}{{\ul{0.1692}\std{0.0064}}}    & {\ul{0.1843}\std{0.0059}}                         & \multicolumn{1}{c}{{{0.0902}\std{0.0037}}}    & \multicolumn{1}{c}{{{0.1448}\std{0.0076}}}    & 0.0468\std{0.0024} & \ul{0.0604}\std{0.0033}                        \\ 
\hline
\textbf{\modelname}                                      & \multicolumn{1}{c}{\textbf{0.3576}\std{0.0007}} & \multicolumn{1}{c}{\textbf{0.3931}\std{0.0014}} & \multicolumn{1}{c}{\textbf{0.2931}\std{0.0010}} & \multicolumn{1}{c|}{\textbf{0.3021}\std{0.0008}} & \multicolumn{1}{c}{\textbf{0.1863}\std{0.0013}} & \multicolumn{1}{c}{\textbf{0.2662}\std{0.0026}} & \multicolumn{1}{c}{\textbf{0.1073}\std{0.0007}} & \textbf{0.1273}\std{0.0008} \\ 
\hline
\end{tabular}
\end{table*}

\textbf{Dataset.} 
We conduct experiments on two real-world datasets: Gowalla-Food
\cite{liu2014exploiting} and \industry. 
Gowalla-Food is a location-based service network for recommendation research.
The \industry dataset is collected from one of the largest local-life service platforms in China. 
Each interaction contains a user-ID, item-ID, timestamp, and GPS locations. 
For each dataset, we use the 10-core settings in \cite{he2020lightgcn} and chronologically split for train, validation, test in 8:1:1.
This means we use the most recent 10\% interactions for testing, which can be regarded as a multi-step sequential recommendation task.
The statistics of all datasets are summarized in Table \ref{tab:datasets}.

\noindent \textbf{Baselines.} 
We compare our \modelname with baselines in two categories (Table \ref{tab:compare}), including:
(1) sequential-based methods, where ST-RNN \cite{liu2016predicting}, STAN \cite{luo2021stan}, TiSASRec \cite{li2020time}, and SLRC \cite{wang2019modeling} utilize spatial or temporal information to capture the dynamic preferences of users;
(2) graph-based methods, where LightGCN \cite{he2020lightgcn}, SAE-NAD \cite{ma2018point}, TGSRec \cite{fan2021continuoustime}, and TGN \cite{rossi2020temporal} capture the collaborative signal and utilize the temporal information from multi-hop neighbors.

\noindent \textbf{Implementation and Evaluation.} 
We implement our \modelname with PyTorch \cite{paszke2019pytorch} and conduct experiments on Tesla V100 (32GB). 
We search the node dimension $c$ and time dimension $c_t$ from \{64, 128, 256\}.
The learning rate is selected from \{1e-3, 1e-4, 1e-5\}, and the layer number $L$ is selected from \{1, 2, 3\}.
For recent neighbor sampling number $M$, we search from \{10, 20, 30\}.
We fix function $f(\cdot)$ as identity and search $\tau$ from \{1, 2, 5\}.
The best version is $c = c_t = 128$, lr = 1e-4, $L = 2$, $M = 20$, $\tau = 1$.
For interaction in the test set, we perform a full ranking \cite{he2020lightgcn} with all item candidates to evaluate the top-K recommendation performance, including Hit@K and NDCG@K, where K in \{10, 20\}.

\begin{figure}[t]
    \centering
    \includegraphics[width=\linewidth]{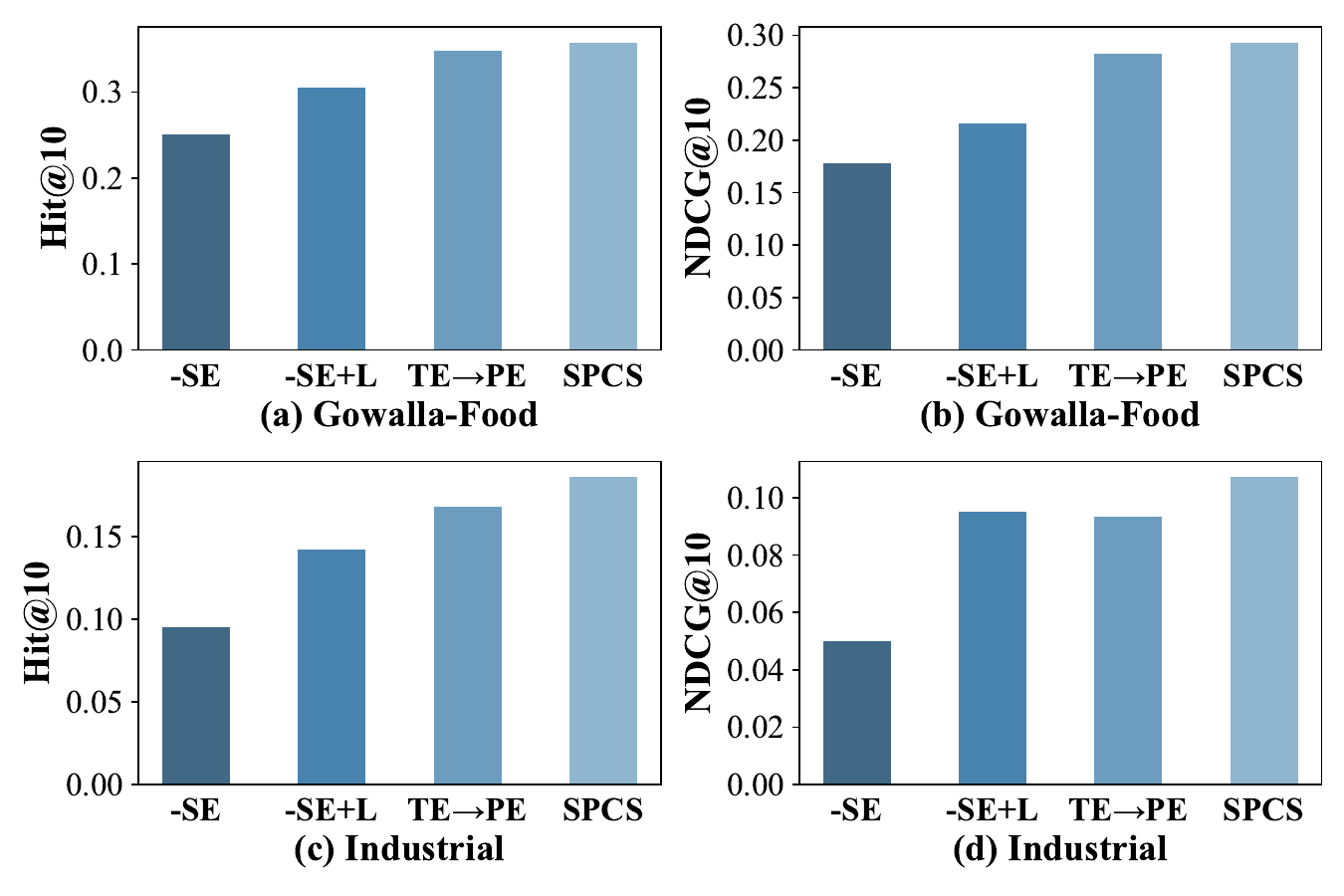}
    \vspace{-0.25in}
    \caption{Ablation study for \modelname on two datasets.}
    \label{fig:ablation}
    \vspace{-0.15in}
\end{figure}

\subsection{Overall Performance}

Table \ref{tab:all} shows the performance comparison between our \modelname and baselines. 
The observations from the table are: 
(1) \modelname consistently outperforms all the baselines on two datasets. 
In particular, \modelname improves over the strongest baseline \textit{w.r.t} Hit@10 by 0.1132 and 0.0961 on Gowalla-Food and \industry datasets, respectively. 
The superiority of our \modelname lies in simultaneously modeling spatiotemporal periodicity with the encoding layer and spatiotemporal graph transformer layer. 
(2) In sequential-based methods, STAN performs best on Gowalla-Food dataset, while SLRC performs best on \industry dataset.
The reason is that SLRC adopts the Hawkes process to explicitly model the temporal periodicity on \industry dataset.
However, they ignore the spatiotemporal collaborative signal so that they perform worse than our \modelname.
(3) TGN performs best in graph-based methods.
This is because TGN captures collaborative signals with temporal information in dynamic graphs.
However, it ignores the spatiotemporal periodicity, especially the spatial information, which makes it worse than our \modelname.

\begin{figure}[tp]
\centering
\subfigure[TGN.]{
\includegraphics[width=0.4\linewidth]{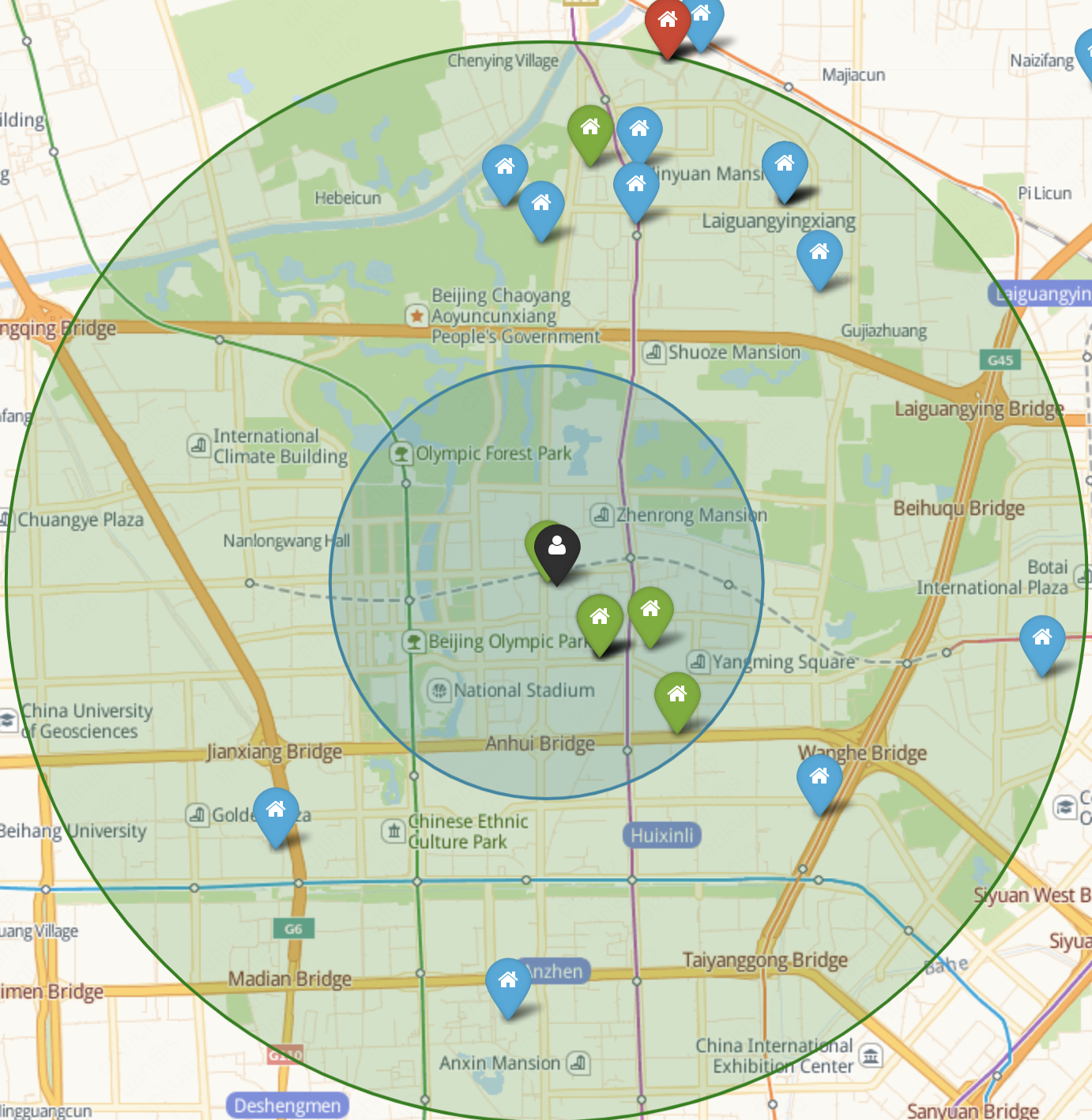}
}
\subfigure[\modelname.]{
\includegraphics[width=0.4\linewidth]{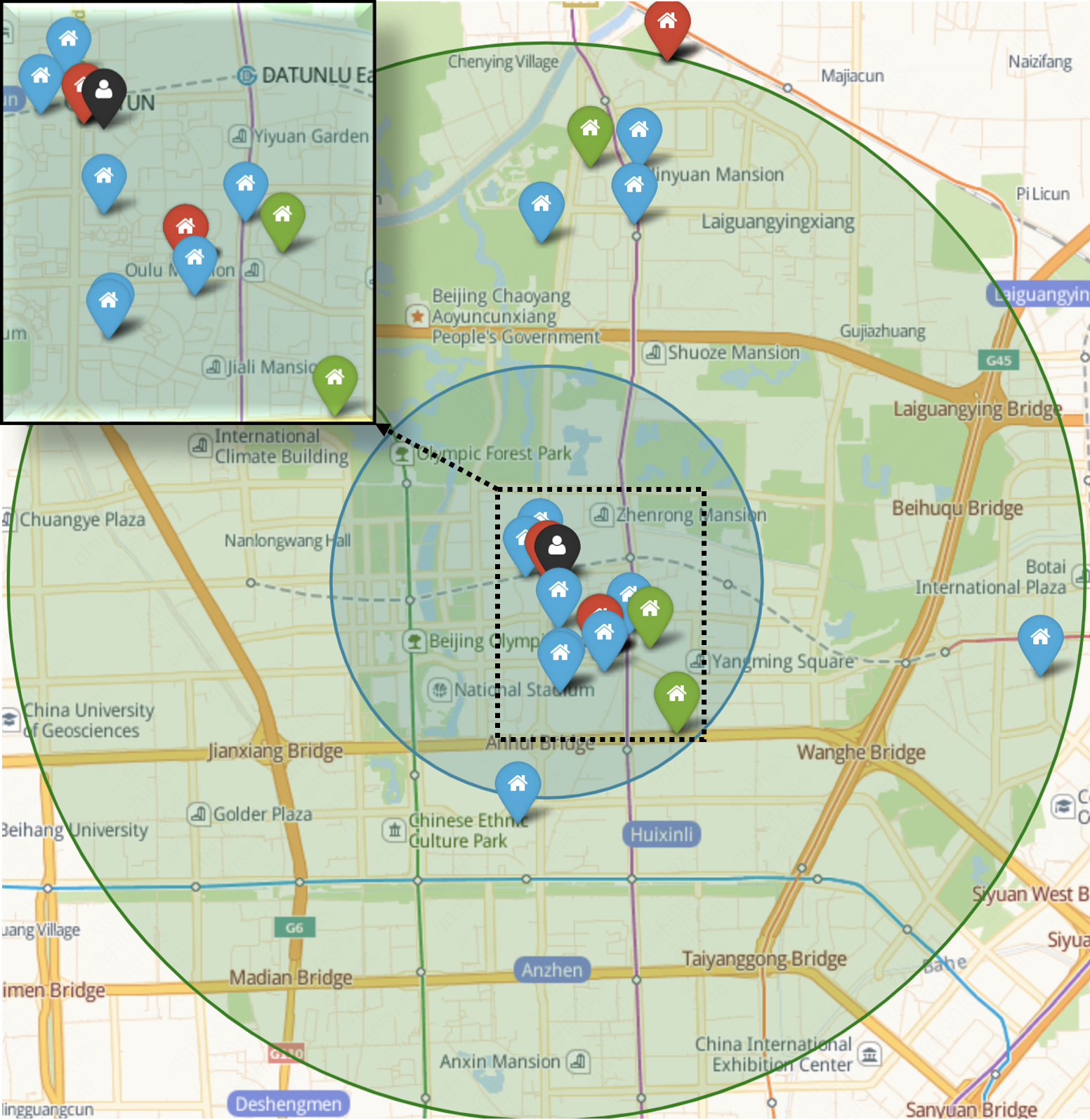}
}
\vspace{-0.1in}
\caption{Case study for user-994 over a period of time on \industry dataset.}
\label{fig:contribution}
\vspace{-0.1in}
\end{figure}

\subsection{Model Analysis of \modelname}

\subsubsection{Ablation Study}

We further conduct an ablation study with several variants to validate the contributions of two encodings in \modelname:
(1) \textit{-SE}, in which spatial encoding is not used;
(2) \textit{-SE+L}, in which spatial encoding is not used but the recall is based on location within 5km;
(3) \textit{TE$\rightarrow$PE}, which replaces temporal encoding with position encoding.
Table \ref{fig:ablation} reports the performance of these variants on two datasets.
Here, we can make the following observations:
(1) variant \textit{-SE} suffers severe performance degradation on two datasets, which demonstrates the importance of spatial encoding in local-life service recommendation.
(2) variant \textit{-SE+L} improves the performance of variant \textit{-SE}, but still performs worse than \modelname.
This indicates the importance of spatial information, and propagating this information in graphs leads to better performance.
(3) The performance degradation of variant \textit{TE$\rightarrow$PE} also demonstrates the crucial role of temporal encoding.

\subsubsection{Case Study}
 
To further investigate the effectiveness of our \modelname, we conduct a case study comparing \modelname with TGN on \industry dataset. 
As shown in Figure \ref{fig:contribution}, the black mark denotes the user’s latest visited location, green marks denote the user’s target items at different times, blue marks denote the top-5 predicted items, and red marks denote the hit items.
The inner blue circle and outer green circle denote a 2km and 5km radius, respectively.
From the results, we find that the user's target items are mostly located within 2km, and \modelname performs better than TGN in hitting these items.
This further demonstrates the importance of simultaneously modeling the spatiotemporal periodicity and collaborative signal for local-life service recommendation.

%% file: 01tex/060conclusion.tex
In this paper, to simultaneously model the spatiotemporal periodicity and collaborative signal for local-life service recommendation, we propose a novel \modelname. 
To capture the spatiotemporal contexts in graphs, we design the encoding layer with temporal encoding and spatial encoding. 
To further capture the spatiotemporal periodicity and collaborative signal, we design a
novel spatiotemporal graph transformer (SGT) layer, which aggregates the spatiotemporal information from multi-hop neighbors.
Extensive experiments on both real-world public and industrial datasets demonstrate the effectiveness of our proposed \modelname. 